\documentclass[aps,pra,showpacs,twocolumn,superscriptaddress, floatfix]{revtex4-1}

\usepackage{graphicx}
\usepackage{amsmath}
\usepackage[utf8]{inputenc}

\usepackage{epsfig}
\usepackage{url}
\usepackage{multirow}
\usepackage{capt-of} 
\usepackage{amsmath}
\usepackage{amssymb}
\usepackage{graphicx}
\usepackage{tikz}
\usepackage{verbatim}
\usepackage{mathrsfs} 
\usepackage{float}
\usepackage{bm}
\usepackage{physics}
\usepackage{mathtools}

\usepackage{algorithm}
\usepackage{algorithmic}
\usepackage{array}
\usepackage{setspace}
\usepackage{ulem}

\begin{document}


\title{Constraint-Aware Quantum Optimization via Hamming Weight Operators}

\author{Yajie Hao}
\affiliation{Institute of Fundamental and Frontier Sciences, University of Electronic Science and Technology of China, Chengdu 611731, China}
\affiliation{Key Laboratory of Quantum Physics and Photonic Quantum Information, Ministry of Education, \\
University of Electronic Science and Technology of China, Chengdu 611731, China}

\author{Qiming Ding}
\email{dqiming94@pku.edu.cn}
\affiliation{Center on Frontiers of Computing Studies, Peking University, Beĳing 100871, China}%
\affiliation{School of Computer Science, Peking University, Beĳing 100871, China}%

\author{Xiao Yuan}
\email{xiaoyuan@pku.edu.cn}
\affiliation{Center on Frontiers of Computing Studies, Peking University, Beĳing 100871, China}%
\affiliation{School of Computer Science, Peking University, Beĳing 100871, China}%

\author{Xiaoting Wang}
\email{xiaoting@uestc.edu.cn}
\affiliation{Institute of Fundamental and Frontier Sciences, University of Electronic Science and Technology of China, Chengdu 611731, China}
\affiliation{Key Laboratory of Quantum Physics and Photonic Quantum Information, Ministry of Education, \\
University of Electronic Science and Technology of China, Chengdu 611731, China}

\begin{abstract}
Constrained combinatorial optimization with strict linear constraints underpins applications in drug discovery, power grids, logistics, and finance, yet remains computationally demanding for classical algorithms, especially at large scales. The Quantum Approximate Optimization Algorithm (QAOA) offers a promising quantum framework, but conventional penalty-based formulations distort optimization landscapes and demand deep circuits, undermining scalability on near-term hardware. In this work, we introduce Hamming Weight Operators, a new class of constraint-aware operators that confine quantum evolution strictly within the feasible subspace. Building on this idea, we develop Adaptive Hamming Weight Operator QAOA, which dynamically selects the most effective operators to construct shallow, problem-tailored circuits. We validate our approach on benchmark tasks from both finance and high-energy physics, specifically portfolio optimization and two-jet clustering with energy balance. Across these problems, our method inherently satisfies all constraints by construction, converges faster, and achieves higher Approximation Ratios than penalty-based QAOA, while requiring roughly half as many gates. By embedding constraint-aware operators into an adaptive variational framework, our approach establishes a scalable and hardware-efficient pathway for solving practical constrained optimization problems on near-term quantum devices.
\end{abstract}

\maketitle
\section{Introduction}

Combinatorial optimization, which seeks binary assignments that minimize or maximize an objective function, lies at the core of critical applications including portfolio optimization, facility location, project scheduling, political districting, energy systems, routing, logistics, and finance. Yet such problems are typically NP-hard, requiring exponential classical resources for exact solutions~\cite{zheng2024solving,heydaribeni2024distributed,jin2025fixed,peres2021combinatorial,banks2008review,huang2024quantum,liu2025solving,lin2025multi,farahani2009facility,dugovsija2020new,kochenberger2004unified,cui2024quantum}. Quantum computing has already shown promise for tackling such challenges across a range of optimization problems~\cite{ma2025experimental,nguyen2023quantum,dlaska2022quantum,yang2025selection,situ2020quantum,zong2024determination,zheng2025quantum,liang2022quantum,wei2023quantum,yan2024universal,montanaro2024quantum,song2024quantum}.
Among the most prominent approaches, the Quantum  Approximate Optimization Algorithm (QAOA) has emerged as a leading framework for combinatorial optimization~\cite{chalupnik2022augmenting,wurtz2022counterdiabaticity,yu2022quantum,zhu2022adaptive,bravyi2020obstacles,hadfield2019quantum,bartschi2020grover,golden2021threshold,fuchs2022constraint,egger2021warm,magann2022feedback,magann2022lyapunov,yoshioka2023fermionic,wurtz2021classically,villalba2021improvement}. By mapping combinatorial optimization problems to Ising Hamiltonians and employing a variational quantum-classical loop, QAOA exploits quantum superposition and interference to explore vast solution spaces in parallel, balancing accuracy and feasibility through circuit depth~\cite{harrigan2021quantum,cerezo2021variational,farhi2014quantum,zhou2020quantum,gao2021quantum,xu2024quafu,ding2024molecular,cheng2024quantum,vcepaite2025quantum,ni2024adaptive,liu2025efficient}. While large-scale quantum advantage is not yet proven, QAOA has already shown performance comparable to or surpassing leading classical heuristics on certain problems, making it a compelling candidate for realizing practical quantum optimization in the near term~\cite{farhi2016quantum,boulebnane2025evidence,shaydulin2024evidence,montanez2025toward,diez2023quantum,blekos2024review,tsvelikhovskiy2025provable,lu2025evidence,wang2025performance,omanakuttan2025threshold,kazi2024analyzing,li2025quantum,ramezani2024reducing,song2023trainability,binkowski2024elementary}.

However, applying QAOA to constrained problems encounters two major bottlenecks. First, the conventional approach to imposing linear constraints---adding penalty terms to the cost Hamiltonian---distorts the optimization landscape, making performance highly sensitive to initialization and hindering reliable convergence~\cite{ayodele2022penalty,brandhofer2022benchmarking,verma2022penalty}. The penalty factor must be delicately tuned: small values fail to ensure feasibility, while large values create steep, rugged energy landscapes that may induce barren plateaus and severely hinder convergence~\cite{mirkarimi2024quantum,herman2023constrained,coello2000use}. Second, the circuit depth required by QAOA grows rapidly with problem size, amplifying the effects of quantum noise and limiting near-term applicability~\cite{willsch2020benchmarking,zhu2022adaptive,chandarana2022digitized,bravyi2020obstacles,xiang2025choco}. This challenge is especially pronounced in constrained settings, since the penalty Hamiltonian often translates into a deep, resource-intensive circuit that further increases gate counts. A scalable and resource-efficient alternative is therefore essential to unlock the full potential of QAOA on noisy intermediate-scale quantum (NISQ) devices~\cite{guo2024experimental,ebadi2022quantum,ma2025experimental, huang2023near,preskill2018quantum,chen2023complexity,wang2021noise}.

\begin{figure*}
\includegraphics[width=1.0\textwidth]{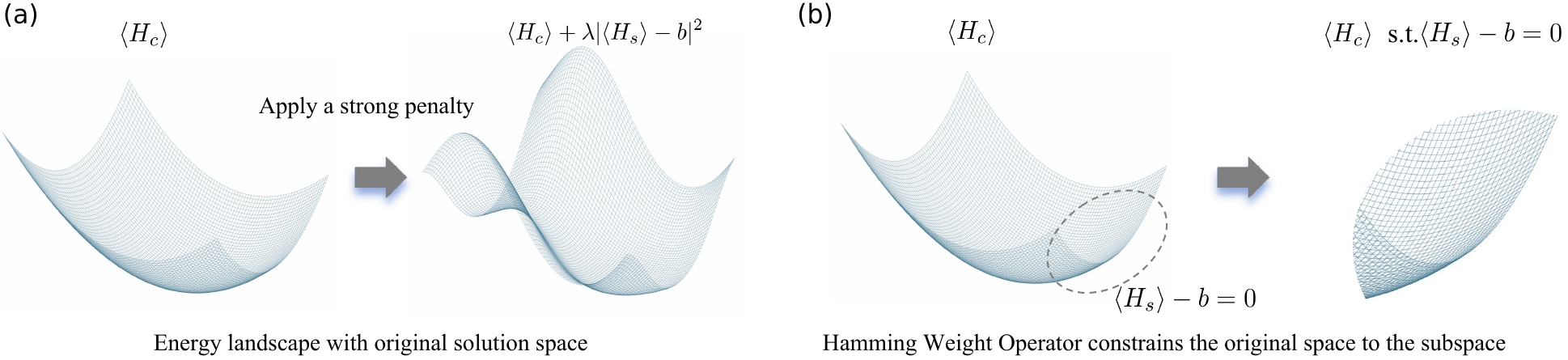}
\caption{Comparison between penalty-based and Hamming Weight Operator approaches for enforcing constraints in QAOA. (a) Penalty-based QAOA distorts the energy landscape with steep barriers. (b) The Hamming Weight Operator directly restricts evolution to the feasible subspace, preserving stability and efficiency. }
\label{fig:fig_1}
\end{figure*}

To address these challenges, we introduce a fundamentally different paradigm: instead of penalizing infeasible states, we design a quantum evolution restricted entirely to the feasible subspace. This is achieved through a novel class of operators, the Hamming Weight Operators, which act as constraint-aware mixers that connect only valid solutions. Building on this foundation, we propose the Adaptive Hamming Weight Operator QAOA (AHWO-QAOA). Inspired by the ADAPT-VQE framework~\cite{grimsley2019adaptive,tang2021qubit}, AHWO-QAOA iteratively constructs a shallow, problem-tailored \textit{ansatz} by adaptively selecting the most effective Hamming Weight Operators from a predefined pool, thereby ensuring both constraint satisfaction and resource efficiency.

We validate AHWO-QAOA through extensive numerical simulations on benchmark portfolio-optimization problems of up to 20 qubits, as well as on the two-jet clustering with energy balance task from high-energy physics. Our results demonstrate three decisive advantages over conventional penalty-based approaches: (i) \textit{guaranteed feasibility}, as all linear constraints are satisfied by construction; (ii) \textit{accelerated convergence}, with high-quality solutions obtained in significantly fewer iterations; and (iii) \textit{reduced resource requirements}, with AHWO-QAOA using nearly half as many elementary gates while achieving higher Approximation Ratios. By validating its performance across both financial and physics-inspired benchmarks, AHWO-QAOA provides a framework that is simultaneously constraint-aware, fast-converging, and hardware-efficient, thereby establishing a necessary foundation for scalable constrained optimization on near-term quantum devices and paving the way toward practical demonstrations of quantum advantage in diverse real-world applications.

\section{Combinatorial Optimization with Linear Constraints}

Combinatorial optimization problems constitute a central class of computational challenges where the decision variables are binary ($x_i \in \{0,1\}$) and must satisfy specific constraints. These problems naturally arise in many practical settings such as finance, logistics, and telecommunications, where discrete decisions are coupled with limited resources. Formally, the problem can be expressed as
\begin{equation}
\label{eq:co_dc}
\begin{split}
\min_{\mathbf{x}} \;& f(\mathbf{x}) = \sum_{i,j} \mu_{ij} x_i x_j + \sum_k \eta_k x_k, \\
\text{s.t. } & \sum_i \omega_i x_i = b.
\end{split}
\end{equation}
where $\mathbf{x} = \{x_1, x_2, \ldots, x_n\}$ are binary decision variables, $\mu_{ij}$ represents pairwise interaction terms (e.g., correlations or conflicts), $\eta_k$ encodes single-variable contributions (e.g., costs or rewards), $\omega_i$ are integer weights, and $b$ is a fixed integer constant.

Linear constraints of this form are ubiquitous in practice and arise across diverse domains. In finance, for instance, portfolio optimization requires selecting a subset of assets under a fixed budget, where $x_i$ indicates whether asset $i$ is included, $\omega_i$ denotes its cost, and $b$ is the total budget~\cite{gunjan2023brief,rebentrost2024quantum,benati2007mixed}. In logistics and scheduling problems such as vehicle routing or job-shop scheduling, $\omega_i$ can represent required capacity or processing time, with $b$ corresponding to the total available resource~\cite{herroelen2005project,azad2022solving,harwood2021formulating,omu2013distributed}. In telecommunications and network design, channel or frequency allocation problems impose bandwidth constraints, where $\omega_i$ reflects bandwidth consumption and $b$ the overall spectrum capacity~\cite{dixit2023quantum,golestan2023quantum,vlachogiannis2008quantum}. Similarly, in manufacturing systems or cloud computing, $\omega_i$ may denote the resource usage of a task, and $b$ the overall system capacity~\cite{jain2023quantum,boulebnane2024solving}.  

Such problems are NP-hard in general and pose significant challenges for classical algorithms, particularly as both the problem size and the number of constraints increase~\cite{xiang2025choco,yan2024universal}. This motivates the exploration of quantum algorithms, such as the QAOA, as a potential approach to tackle constrained combinatorial optimization problems more efficiently.

To solve this constrained combinatorial optimization problem in Eq.~\eqref{eq:co_dc} using QAOA, we first map the classical binary variables to quantum operators via $x_i \mapsto (\mathbf{1}-\sigma_z^i)/2$, where $\sigma_z^i$ is the Pauli-$Z$ operator acting on qubit $i$. Under this mapping, the cost function is represented by the cost Hamiltonian
\begin{equation}
\label{eq:Hc}
H_c = \frac{1}{4}\sum_{i,j} \mu_{ij}(\mathbf{1}-\sigma_z^i)(\mathbf{1}-\sigma_z^j)
+ \frac{1}{2}\sum_k \eta_k (\mathbf{1}-\sigma_z^k),
\end{equation}
while the linear constraint is encoded as
\begin{equation}
\label{eq:Hs}
H_s = \sum_i \omega_i \frac{\mathbf{1}-\sigma_z^i}{2}.
\end{equation}
 
The goal is to prepare a variational quantum state $\ket{\psi(\boldsymbol{\gamma}, \boldsymbol{\beta})}$ using QAOA, such that
\begin{equation}
\begin{split}
\min &\;\bra{\psi(\boldsymbol{\gamma}, \boldsymbol{\beta})} H_c \ket{\psi(\boldsymbol{\gamma}, \boldsymbol{\beta})}, \\
\text{s.t. } &\;\bra{\psi(\boldsymbol{\gamma}, \boldsymbol{\beta})} H_s \ket{\psi(\boldsymbol{\gamma}, \boldsymbol{\beta})} = b.
\end{split}
\end{equation}
Where the variational state with $p$ layers is
\begin{equation}
\label{eq:qaoa_ansatz}
\ket{\psi(\boldsymbol{\gamma}, \boldsymbol{\beta})} =
\prod_{l=1}^p e^{-i\beta_l H_m}e^{-i\gamma_l H_c}\ket{\psi_0},
\end{equation}
parameterized by the angles $\boldsymbol{\gamma} = \{\gamma_1,\ldots,\gamma_p\}$ and $\boldsymbol{\beta} = \{\beta_1,\ldots,\beta_p\}$, the initial state $\ket{\psi_0} = \ket{+}^{\otimes n}$. These variational parameters control the evolution times under $H_c$ and $H_m = \sum_{i=1}^n \sigma_x^i$, respectively. An extended variant, known as Multi-Angle or Full-Parameter QAOA (ma-QAOA/FP-QAOA)~\cite{herrman2022multi,shi2022multiangle}, which can effectively reduce the number of layers at the cost of increasing the parameters optimized for each layer. All simulations in this work employ this full-parameter formulation unless otherwise noted.

In practice, the constraint is usually incorporated into the optimization through a penalty method. The resulting augmented loss function is
\begin{equation}
\label{eq:new_loss}
\begin{split}
L(\boldsymbol{\gamma}, \boldsymbol{\beta}) = &
\bra{\psi(\boldsymbol{\gamma}, \boldsymbol{\beta})}H_c\ket{\psi(\boldsymbol{\gamma}, \boldsymbol{\beta})}\\
&+ \lambda |\bra{\psi(\boldsymbol{\gamma}, \boldsymbol{\beta})}H_s\ket{\psi(\boldsymbol{\gamma}, \boldsymbol{\beta})}-b|^2,
\end{split}
\end{equation}
where $\lambda \gg 1$ is a penalty factor chosen to enforce feasibility of the constraint.  

Although conceptually straightforward, this penalty-based formulation has significant drawbacks. A small $\lambda$ may lead to infeasible solutions that violate the constraint, while an excessively large $\lambda$ creates an ill-conditioned optimization landscape that hampers convergence and causes instability, as shown in Fig.~\ref{fig:fig_1}(a). Furthermore, large penalty factors typically require deeper circuits and longer optimization times, thereby increasing the computational cost~\cite{mirkarimi2024quantum,herman2023constrained,coello2000use}. These limitations expose the fragility of penalty-based QAOA and motivate the development of alternative approaches for handling constraints more effectively.

\section{Adaptive Hamming Weight Operator QAOA}
\begin{figure*}[htbp]
\centering
\includegraphics[width=1.0\textwidth]{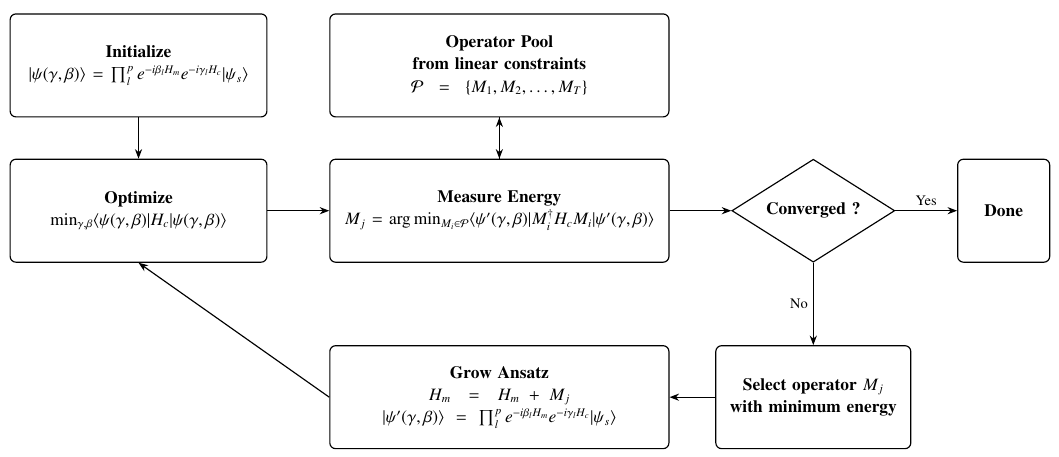}
    \caption{Workflow of the AHWO-QAOA. The algorithm begins by initializing the variational state with the current cost Hamiltonian $H_c$ and mixing Hamiltonian $H_m$. 
An operator pool $\mathcal{P}=\{M_1,M_2,\ldots,M_T\}$ is constructed from the linear constraints. 
At each iteration, the energy contribution of candidate operators is evaluated, and the operator with the largest contribution (lowest energy) is selected. 
The mixing Hamiltonian is then updated by including the chosen operator, and the variational ansatz is grown accordingly. 
This adaptive procedure continues until convergence is reached, thereby reducing circuit depth while preserving constraint satisfaction.} 
    \label{fig:fig_2}
\end{figure*}

In the previous section, we addressed constrained combinatorial optimization using QAOA with a penalty term. While straightforward, this method suffers from instability and inefficiency, as the optimization is highly sensitive to the choice of penalty factor. To overcome these limitations, we propose an alternative strategy: embedding linear constraints directly into the mixing Hamiltonian. This eliminates penalty terms and guarantees that the variational state remains strictly within the feasible subspace, thereby improving both stability and efficiency.

\begin{figure*}
\centering
\includegraphics[width=0.95\textwidth]{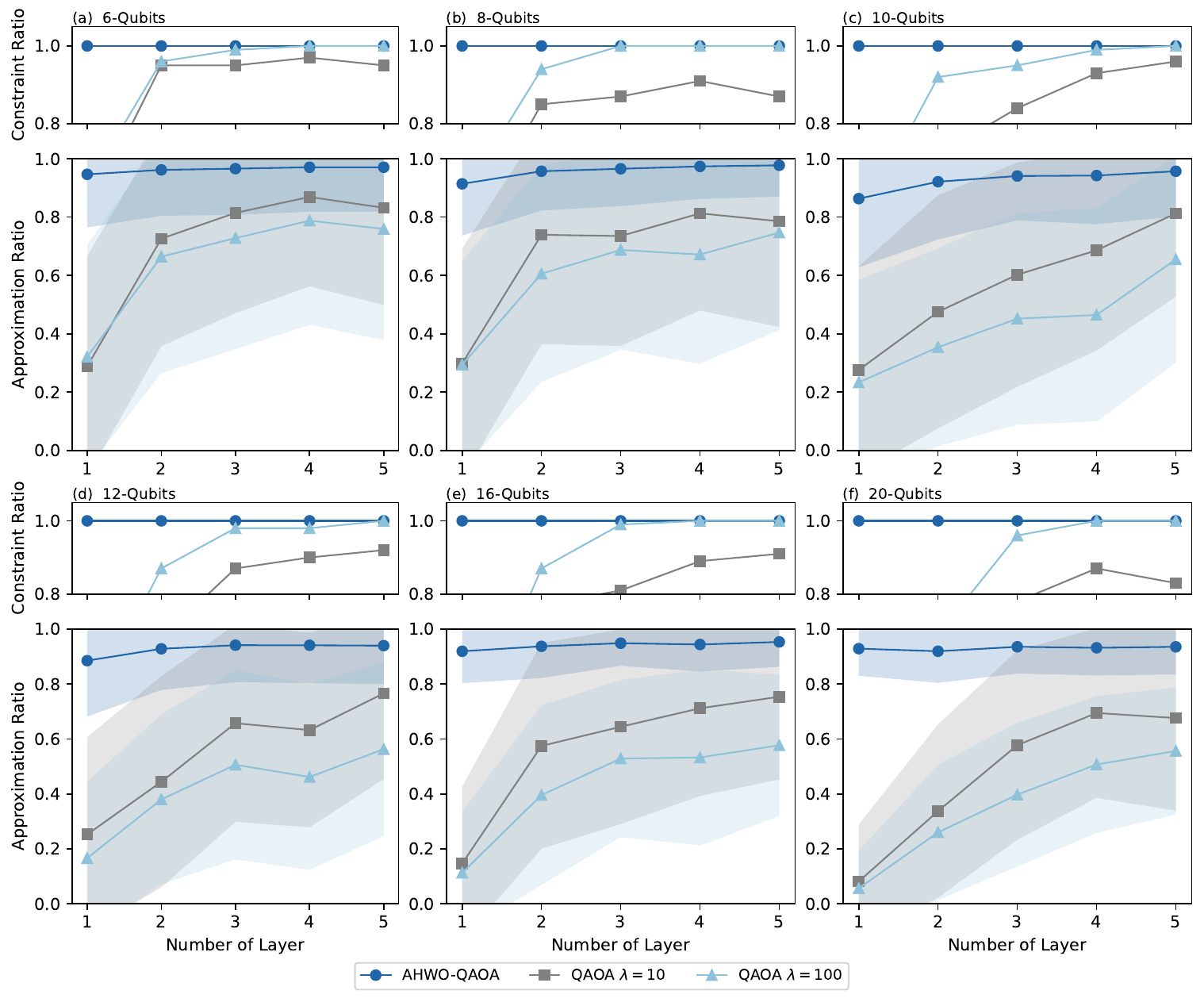}
\caption{Performance comparison between AHWO-QAOA and penalty-based QAOA with penalty factors $\lambda=10$ and $\lambda=100$. 
Each panel corresponds to a system size of (a) 6 qubits, (b) 8 qubits, (c) 10 qubits, (d) 12 qubits, (e) 16 qubits, and (f) 20 qubits. 
The upper plots show the Constraint Ratio, i.e., the percentage of test cases that satisfy the linear constraints, while the lower plots show the Approximation Ratio, defined as $1 - |\langle H_c \rangle - E_0|/|E_0|$ and set to zero whenever $\langle H_s \rangle \neq b$. 
Results are averaged over 100 randomly generated problem instances for each system size. 
Penalty-based QAOA exhibits a trade-off between approximation performance and constraint satisfaction: small $\lambda$ yields higher Approximation Ratios but poor feasibility, while large $\lambda$ enforces constraints at the expense of performance. 
In contrast, AHWO-QAOA consistently satisfies all constraints across all problem sizes and achieves superior Approximation Ratios even with shallow circuits (e.g., one layer with 20 qubits), demonstrating its scalability.}
\label{fig:fig_3}
\end{figure*}

\subsection{Hamming Weight Operator}

We begin with general linear constraints of the form
\begin{equation}
\label{eq:more_con}
\sum_i \omega_i x_i = b, \quad \omega_i \in \mathbb{Z}.
\end{equation}
Such constraints often induce multiple relations among subsets of $\{\omega_i\}$, which we refer to as Hamming Weight Equations:
\begin{equation}
\label{eq:omega_eq}
\sum_{r=1}^{n} \omega_{i_r} = \sum_{k=1}^{m} \omega_{j_k} \leq b,
\end{equation}
where $\{i_r\}$ and $\{j_k\}$ are disjoint index sets.  

As an example, let $\omega_1=1,\,\omega_2=2,\,\omega_3=2,\,\omega_4=3,\,\omega_5=5$, with $b=4$. The induced relations are
\[
\omega_1+\omega_4=\omega_2+\omega_3,\quad 
\omega_1+\omega_2=\omega_4,\quad 
\omega_2=\omega_3,
\]
each giving rise to feasible state exchanges. For instance, from $\omega_1+\omega_4=\omega_2+\omega_3$, both $\ket{1001\ldots}$ and $\ket{0110\ldots}$ are feasible. Similarly, for $\omega_1 + \omega_2 = \omega_4$, if $\ket{11\ldots 0_4 \ldots}$ is feasible, then so is $\ket{00\ldots 1_4 \ldots}$. This invariance under exchanges is what we call the Hamming Weight Invariance.  

To exploit this property, we introduce the Hamming Weight Operator. For Eq.~\eqref{eq:omega_eq}, it is defined as
\begin{equation}
\begin{split}
\label{eq:hamming_op}
M = & \prod_{r=1}^n \prod_{k=1}^m 
\left(\sigma_x^{i_r}+i\sigma_y^{i_r}\right)
\left(\sigma_x^{j_k}-i\sigma_y^{j_k}\right) \\
&+ \prod_{r=1}^n \prod_{k=1}^m 
\left(\sigma_x^{i_r}-i\sigma_y^{i_r}\right)
\left(\sigma_x^{j_k}+i\sigma_y^{j_k}\right),
\end{split}
\end{equation}
whose action is

\begin{equation}
\begin{split}
\label{eq:hamming_in}
M\ket{...0_{i_1}...0_{i_n}...1_{j_1}...1_{j_m}} &= \ket{...1_{i_1}...1_{i_n}...0_{j_1}...0_{j_m}}, \\
M\ket{...1_{i_1}...1_{i_n}...0_{j_1}...0_{j_m}} &= \ket{...0_{i_1}...0_{i_n}...1_{j_1}...1_{j_m}}, \\
M\ket{\text{others}} &= 0.
\end{split}
\end{equation}
Thus $M$ exchanges feasible basis states while annihilating infeasible ones, ensuring dynamics remain confined to the constraint subspace.  For any feasible state $\ket{\psi_{s}}$ that satisfies the constraints, the transformed state
\begin{equation}
e^{-i\sum M}\ket{\psi_{s}},
\end{equation}
remains a feasible state, thereby enabling exploration of the constrained subspace without violating the linear constraints.

The operator is a generalized swap-like construction. In the two-qubit case with equal weights, Eq.~\eqref{eq:hamming_op} reduces to the familiar form $\sigma_x^p\sigma_x^q + \sigma_y^p\sigma_y^q$~\cite{hadfield2019quantum, robledo2025chemistry}. Hence, the Hamming Weight Operator generalizes the XX+YY operator to higher dimensions.

\subsection{Hamming Weight Operator QAOA}

Based on the above, QAOA can be reformulated to incorporate Hamming Weight Operators. For the constrained optimization problem in Eq.~\eqref{eq:co_dc}, we first enumerate all induced Hamming Weight Equations and construct the corresponding operators $M_t$. The modified mixing Hamiltonian and variational state are then
\begin{equation}
\label{eq:ham_op_qaoa}
\begin{split}
H_m &= \sum_{i \notin \mathbf{c}} \sigma_x^i + \sum_t M_t, \\
\ket{\psi(\boldsymbol{\gamma}, \boldsymbol{\beta})} &= 
\prod_{l=1}^p e^{-i\beta_l H_m} e^{-i\gamma_l H_c}\ket{\psi_s},
\end{split}
\end{equation}
where $\mathbf{c}$ is the set of constrained qubits. By construction, $\ket{\psi(\boldsymbol{\gamma}, \boldsymbol{\beta})}$ remains in the feasible subspace.  

The optimization objective is then
\begin{equation}
\label{eq:expect_over}
F(\boldsymbol{\gamma}, \boldsymbol{\beta}) =
\bra{\psi(\boldsymbol{\gamma}, \boldsymbol{\beta})} H_c \ket{\psi(\boldsymbol{\gamma}, \boldsymbol{\beta})}.
\end{equation}
This approach encodes linear constraints directly into the ansatz, avoiding penalty terms and improving stability relative to Eq.~\eqref{eq:new_loss}.

\subsection{Adaptive Hamming Weight Operator QAOA}

Although Hamming Weight Operator QAOA ensures feasibility, circuit depth may grow substantially as the number of equations increases, since each adds a nontrivial operator $M_t$. To address this, we propose the Adaptive Hamming Weight Operator QAOA (AHWO-QAOA).  

Inspired by the ADAPT-VQE, we build an operator pool $\mathcal{P}=\{M_1,M_2,\ldots,M_T\}$ from all candidate operators. During training, operators are adaptively selected based on their contribution to lowering the cost Hamiltonian. At each iteration, candidate operators are evaluated based on the resulting energy, and the operator that yields the lowest energy is selected and included in the ansatz. This process repeats until convergence, as shown in Fig.~\ref{fig:fig_2}. 

In the idealized construction of a constraint-preserving mixer, one might attempt to enumerate all induced Hamming Weight Equations so that any pair of constrained qubits could, in principle, be coupled through at least one operator. However, the total number of such equations increases quickly with the problem size $n$, making a full enumeration difficult to handle in practice and leading to an overly large and inefficient operator pool.
To avoid this, in AHWO-QAOA we do not attempt to list all possible equations. Instead, we adopt a more efficient strategy in which only a sparse, connectivity-preserving subset of Hamming Weight Operators is constructed. The key idea is to ensure that every constrained qubit participates in at least one operator and that these operators collectively form a chain that covers all constraint weights. This guarantees that the feasible Hamming-weight space remains connected under the action of the mixer, while requiring only $O(n)$ operators in total. Importantly, this operator set can be obtained efficiently using a polynomial-time prefix-based search with worst-case complexity $O(n^2)$, whose detailed construction and analysis are provided in Appendix~A.

As a result, AHWO-QAOA dynamically balances expressibility and resource efficiency. By pruning redundant operators, it achieves shallower circuits while preserving constraint satisfaction and often improving optimization performance, making it particularly well-suited for near-term quantum devices.

\subsection{Compatibility with Other QAOA Variants}

Although the numerical experiments in this work adopt the Multi-Angle QAOA~\cite{herrman2022multi,shi2022multiangle} as the primary baseline, it is important to emphasize that a broad family of enhanced QAOA variants has been developed in recent years to improve convergence speed, avoid barren plateaus, or reduce circuit complexity. Representative examples include DC-QAOA~\cite{chandarana2022digitized,wurtz2022counterdiabaticity}, GM-QAOA~\cite{bartschi2020grover}, Quantum Dropout~\cite{wang2023quantum}, WS-QAOA~\cite{egger2021warm}, Recursive QAOA~\cite{bravyi2020obstacles}, as well as several other variants proposed in the recent literature~\cite{blekos2024review}.

These variants modify different components of the alternating-operator framework—such as parameter schedules, measurement strategies, layer-wise coupling structures, or gradient estimation—while still relying on the standard division between a cost Hamiltonian and a mixing Hamiltonian. In contrast, our AHWO-QAOA introduces a modification exclusively to the \emph{mixer}, replacing single-qubit $X$ operations with constraint-preserving Hamming Weight Operators.

Because AHWO-QAOA modifies only the structure of the mixer and leaves the cost Hamiltonian and variational training protocol unchanged, it is orthogonal and fully compatible with the variants listed above. That is, the constraint-preserving mixer proposed in this work may, in principle, be combined with other QAOA improvements such as dynamic couplings, warm starts, recursive structures, or dropout-based training. In this sense, AHWO-QAOA provides a modular upgrade that enforces feasibility by construction and can serve as a drop-in replacement for the standard mixer in a wide class of QAOA frameworks.

\section{Application}\label{sec:7}
\subsection{Portfolio Optimization}

A representative instance of the constrained combinatorial optimization problem in Eq.~\eqref{eq:co_dc} arises in the context of portfolio optimization in financial markets. Here, an investor must select a subset of assets from a candidate pool in order to minimize investment risk or maximize expected return, subject to a strict budget constraint.  

We define binary decision variables $\{x_i\}$ ($i = 1, 2, \ldots, n$), where $x_i = 1$ indicates that asset $i$ is included in the portfolio and $x_i = 0$ otherwise. The quadratic coefficients $\mu_{ij}$ capture the pairwise correlation between assets $i$ and $j$, reflecting diversification effects or joint risk contributions. The linear coefficients $\eta_k$ encode the expected return or individual risk of selecting asset $k$. The objective function therefore balances portfolio risk and return:  
\[
f(\mathbf{x}) = \sum_{i,j} \mu_{ij} x_i x_j + \sum_k \eta_k x_k.
\]

A strict budget constraint is imposed as
\[
\sum_i \omega_i x_i = b,
\]
where $\omega_i$ denotes the cost (e.g., normalized price) of asset $i$ and $b$ is the available budget. This ensures that the total investment does not exceed the investor’s resources.  

Such formulations are widely used in financial engineering. For instance, in constructing an equity portfolio, $\omega_i$ may correspond to the capital required to purchase one share of stock $i$, while $\eta_k$ is derived from historical expected returns. The quadratic term $\mu_{ij}$ models asset covariances, penalizing highly correlated selections and promoting diversification. The resulting optimization problem is both practically relevant and computationally challenging, as it combines binary decision variables with hard linear constraints. This makes portfolio optimization a realistic benchmark for assessing the effectiveness of the proposed AHWO-QAOA.  

\begin{figure}[H]
\centering
\includegraphics[width=0.45\textwidth]{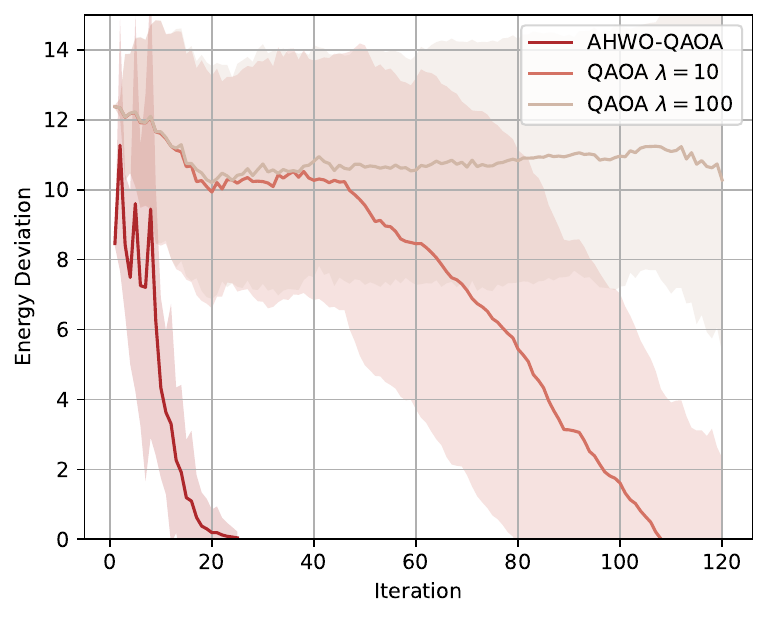}
\caption{Convergence behavior of AHWO-QAOA compared to penalty-based QAOA with penalty factors $\lambda=10$ and $\lambda=100$ for a 12-qubit instance. 
The vertical axis shows the Energy Deviation, defined as $|\langle H_c \rangle - E_0|$, where $\langle H_c \rangle$ is the expectation value of the cost Hamiltonian and $E_0$ its ground state energy. 
Each curve represents the mean over 100 random initializations of the variational parameters, and shaded regions indicate the variance across trials. 
Penalty-based QAOA converges slowly and exhibits large fluctuations, especially in early iterations, reflecting poor stability. In contrast, AHWO-QAOA converges significantly faster, with only minor fluctuations in the initial iterations due to its adaptive mechanism, and rapidly stabilizes to reach convergence in fewer iterations.}
\label{fig:fig_4}
\end{figure}

To evaluate performance, we simulated problem instances of size $n \in \{6, 8, 10, 12, 16, 20\}$ qubits. For each size, $100$ independent problem instances were generated with random coefficients $\mu_{ij}$ and $\eta_k$, along with randomly chosen linear constraints consistent with Eq.~\eqref{eq:co_dc}. The variational parameters $\boldsymbol{\gamma}$ and $\boldsymbol{\beta}$ were initialized uniformly at random in the interval $(-0.1,\,0.1)$.

Constraint satisfaction and approximation quality.
We first compared AHWO-QAOA with penalty-based QAOA using penalty factors $\lambda=10$ and $\lambda=100$. The Approximation Ratio was defined as
\[
\text{Approximation Ratio} = 1 - \frac{|\langle H_c \rangle - E_0|}{|E_0|},
\]
where $E_0$ is the ground-state energy of $H_c$. To emphasize feasibility, the ratio was set to zero whenever $\langle H_s \rangle \neq b$. We also measured the Constraint Ratio, the fraction of test cases that satisfy the constraint.

As shown in Fig.~\ref{fig:fig_3}, penalty-based QAOA exhibits a trade-off: small $\lambda$ yields higher Approximation Ratios but poor constraint satisfaction, while large $\lambda$ enforces feasibility at the expense of approximation quality. In contrast, AHWO-QAOA consistently satisfies all constraints and maintains strong approximation performance across all system sizes, even with a single layer and up to 20 qubits, demonstrating clear scalability.

\begin{figure}[H]
\centering
\includegraphics[width=0.45\textwidth]{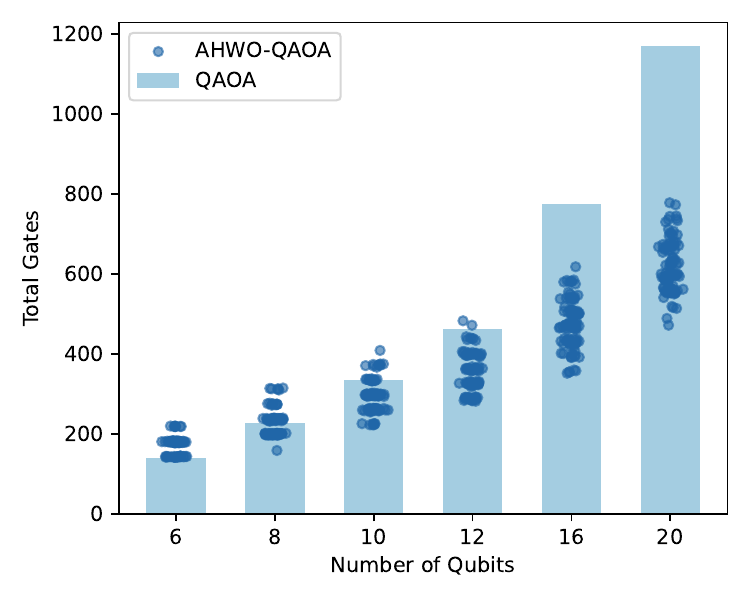}
\caption{Comparison of quantum resource requirements between AHWO-QAOA (1 \textit{ansatz} layer) and penalty-based QAOA (5 \textit{ansatz} layers). 
The vertical axis shows the total number of quantum gates, and the horizontal axis indicates the number of qubits. 
Bars correspond to penalty-based QAOA, while dots represent AHWO-QAOA across 100 randomly generated problem instances. 
The results show that the gate count of penalty-based QAOA grows rapidly with system size, whereas AHWO-QAOA consistently requires fewer gates. 
At 20 qubits, AHWO-QAOA uses approximately half as many gates as penalty-based QAOA while simultaneously achieving superior performance in terms of Approximation Ratio and constraint satisfaction, thus significantly reducing quantum resource costs.}
\label{fig:fig_5}
\end{figure}

Convergence behavior. 
We then analyzed convergence on a representative 12-qubit instance using 100 random initializations. As shown in Fig.~\ref{fig:fig_4}, penalty-based QAOA converges slowly and exhibits large variance, reflecting instability. By contrast, AHWO-QAOA converges significantly faster. Although minor fluctuations occur in the early iterations due to its adaptive mechanism, the method rapidly stabilizes and converges smoothly, requiring far fewer iterations. This demonstrates its superior convergence properties.  

Quantum resource requirements.
Finally, we compared resource costs. From the results above, AHWO-QAOA with only one layer already outperforms penalty-based QAOA with five layers in Approximation Ratio and constraint satisfaction. To provide a fair resource comparison, we therefore evaluated AHWO-QAOA (1 layer) against penalty-based QAOA (5 layers).

As shown in Fig.~\ref{fig:fig_5}, the number of quantum gates for penalty-based QAOA grows rapidly with system size, while AHWO-QAOA consistently requires fewer gates. At 20 qubits, AHWO-QAOA uses roughly half as many gates while still achieving superior performance. This highlights that AHWO-QAOA not only improves accuracy and stability but also dramatically reduces quantum resource costs.

\subsection{Two-Jet Clustering with Energy Balance}

Another representative instance of the constrained combinatorial optimization problem in Eq.~\eqref{eq:co_dc} arises in the context of jet clustering in high-energy physics. Specifically, we consider the two-jet partitioning task, where the goal is to assign final-state particles produced in an $e^+e^-$ collision into two jets consistent with the underlying partonic process (e.g., $H \to s\bar{s}$)\cite{Zhu2025JetClustering,liang2024jet}.  

We define binary decision variables $\{x_i\}$ ($i = 1, 2, \ldots, n$), where $x_i = 1$ indicates that particle $i$ is assigned to jet A and $x_i = 0$ indicates assignment to jet B. The quadratic coefficients $\mu_{ij}$ encode the angular separation between particles $i$ and $j$, reflecting the fact that particles with large opening angles are more likely to originate from different jets. The objective function thus maximizes inter-jet separation, analogous to the Max-Cut formulation:  
\[
f(\mathbf{x}) = \sum_{i,j} \mu_{ij} x_i (1 - x_j).
\]  

In addition to the angular objective, a strict linear constraint is imposed to enforce energy balance between the two jets:  
\[
\sum_i \epsilon_i x_i = \tfrac{1}{2}\sum_i \epsilon_i,
\]
where $\epsilon_i$ denotes the measured energy of particle $i$, provided by experimental data or Monte Carlo event simulations. This constraint ensures that the visible energy carried by jet A equals that of jet B, consistent with the expectation for two-body decays in the center-of-mass frame.

\begin{figure*}
\includegraphics[width=1.0\textwidth]{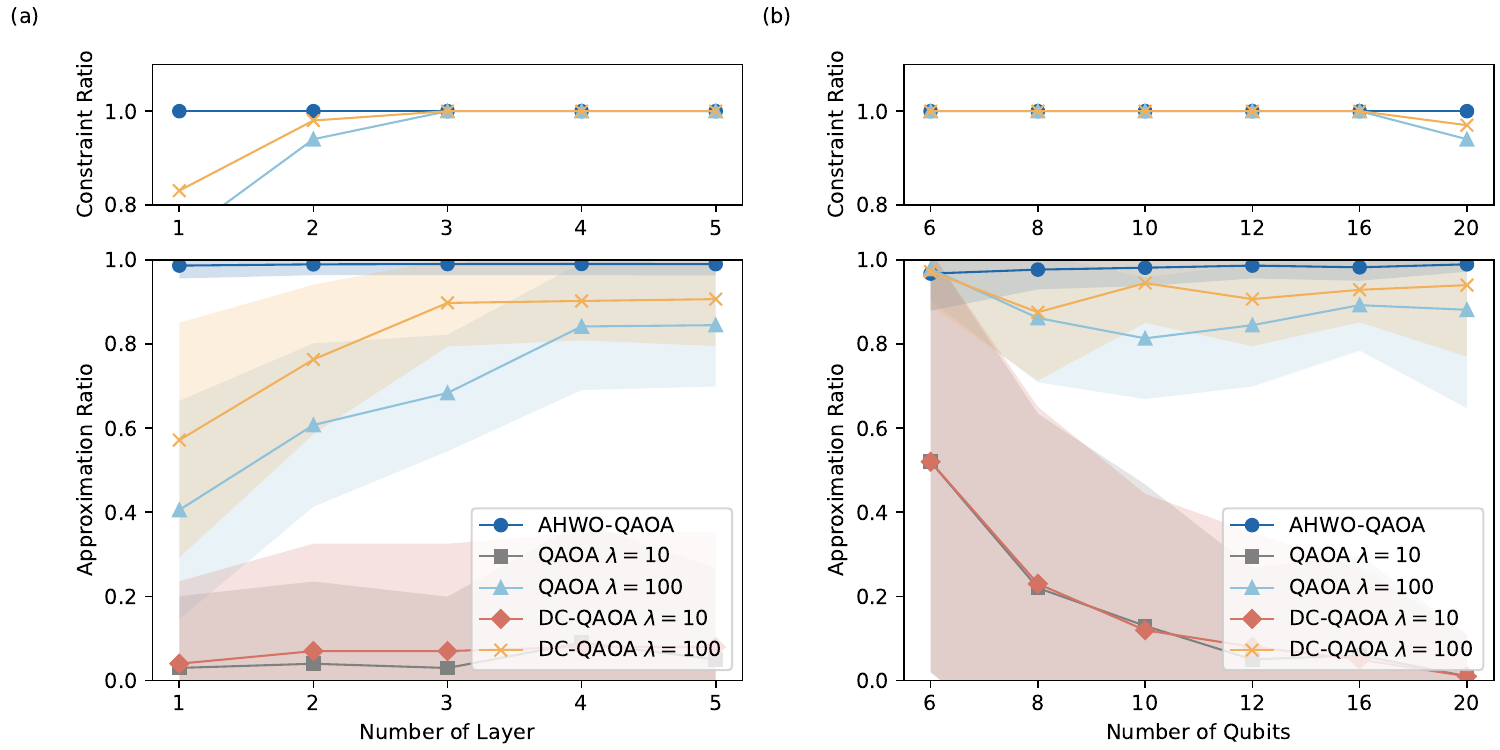}
\caption{
Performance comparison among Adaptive Hamming Weight Operator QAOA (AHWO-QAOA), penalty-based QAOA, and penalty-based DC-QAOA on the Two-Jet Clustering with 
Energy Balance problem. (a) Fixed at 12 qubits, varying the number of ansatz layers from 1 to 5. 
(b) Fixed at 1 layer for AHWO-QAOA and 5 layers for the penalty-based baselines, with qubit numbers varying across $\{6, 8, 10, 12, 16, 20\}$. 
The upper panels report the Constraint Ratio (fraction of solutions satisfying the linear constraint), while the lower panels show the 
Approximation Ratio. Across all settings, AHWO-QAOA consistently satisfies all constraints and 
achieves higher approximation ratios with significantly shallower circuits. The comparison further shows that AHWO-QAOA outperforms both penalty-based QAOA and the enhanced penalty-based DC-QAOA, demonstrating the effectiveness of enforcing linear constraints directly within the mixer.
}

\label{fig:fig_6}
\end{figure*}

This formulation is physically well-motivated. For instance, in $e^+e^- \to ZH$ events with $H \to s\bar{s}$, the Higgs boson decays into two nearly back-to-back partons of comparable energy. The quadratic term $\mu_{ij}$ captures the angular geometry of the final state, while the linear constraint enforces energy conservation and suppresses unphysical solutions where one jet absorbs almost all of the energetic particles. Overall, the problem combines binary jet-assignment variables with a hard linear constraint, making it both practically relevant and computationally challenging, and thus a meaningful benchmark for assessing AHWO-QAOA.  

To further assess the effectiveness of the proposed constraint-preserving mixer, we introduce an additional comparison with a representative enhanced QAOA variant—DC-QAOA—on the two-jet clustering benchmark. Although such variants can in principle be applied to both benchmark problems considered in this work, we select DC-QAOA here because it is one of the enhanced QAOA approaches that has been widely discussed in recent literature. Accordingly, in this subsection we compare penalty-based QAOA, penalty-based DC-QAOA, and our constraint-preserving AHWO-QAOA under identical settings, providing a supplementary baseline that helps illustrate the practical advantages gained by enforcing linear constraints directly within the mixer.

To evaluate performance, we carried out simulations on problem sizes ranging from $n=6$ to $n=20$ qubits. Each instance was generated with random angular coefficients $\mu_{ij}$ and particle energies $\epsilon_i$. As in the portfolio optimization study, we compared AHWO-QAOA against penalty-based QAOA with penalty factors $\lambda=10$ and $\lambda=100$, and we additionally included penalty-based DC-QAOA as a representative enhanced variant recently discussed in the literature. Fig.~\ref{fig:fig_6} summarizes the results. Panel (a) fixes the system size at 12 qubits and varies the number of ansatz layers from 1 to 5, while panel (b) fixes the depth of AHWO-QAOA at 1 layer and the depth of the penalty-based baselines (QAOA and DC-QAOA) at 5 layers to provide a fair comparison across system sizes. The upper panels report the Constraint Ratio, while the lower panels show the Approximation Ratio.

Beyond constraint satisfaction and approximation quality, AHWO-QAOA demonstrates clear advantages in both convergence and resource efficiency. For a 12-qubit instance, single-layer AHWO-QAOA converges in roughly 30 iterations on average, compared to about 90 iterations for five-layer penalty-based QAOA with $\lambda=10$, over 410 iterations with $\lambda=100$, and similarly slower convergence trends observed for penalty-based DC-QAOA. In terms of circuit cost, at 20 qubits AHWO-QAOA requires only about 470 gates, whereas the penalty-based baselines require significantly deeper circuits (e.g., approximately 1270 gates for five-layer QAOA), with DC-QAOA exhibiting a comparable or higher cost due to its additional counterdiabatic terms. These results highlight that AHWO-QAOA not only converges faster and more stably, but also reduces gate counts by more than half relative to the penalty-based approaches.

In summary, AHWO-QAOA consistently satisfies all constraints and achieves near-optimal Approximation Ratios with dramatically shallower circuits, whereas penalty-based QAOA and penalty-based DC-QAOA suffer from a trade-off between feasibility, accuracy, and circuit depth. These advantages become increasingly pronounced as system size grows, underscoring both the scalability and hardware efficiency of AHWO-QAOA for realistic constrained optimization tasks in high-energy physics.

\section{Conclusion}
In this work, we presented the AHWO-QAOA, a framework designed to overcome a fundamental problem in combinatorial optimization with linear constraints: the reliance on resource-intensive and unstable penalty methods for constraint handling. By introducing a new class of constraint-preserving gates, the Hamming Weight Operator, and integrating them into an adaptive ansatz construction, our approach changes the main strategy from punishing infeasible states to a quantum evolution confined entirely within the feasible subspace. Our numerical simulations up to 20 qubits confirm that this approach is not only workable but also highly effective, delivering feasible solutions with accelerated convergence and substantially reduced quantum resources compared to conventional methods.

The significance of our findings extends far beyond this specific implementation of portfolio optimization and jet clustering, offering both theoretical and practical value for the broader field of quantum computing. From a theoretical standpoint, the Hamming Weight Operator is not limited to QAOA; it serves as a new, general-purpose building block for constructing constraint-aware circuits. This tool can be easily adapted for other variational algorithms like VQE, or in quantum machine learning models where preserving specific symmetries or properties is crucial. More broadly, AHWO-QAOA provides a scalable template for tackling other complex constrained problems, showing the importance of designing the quantum dynamics in alignment with the structure of the constraint set. Future directions include extending the framework to inequality constraints, exploring richer symmetries and conservation laws, and benchmarking on real hardware. By delivering an effective, resource-efficient, and penalty-free approach, this work advances the development of practical quantum optimization for a wide range of real-world applications.

\bibliographystyle{apsrev4-1}
\bibliography{ref}

\end{document}